\documentclass[two column,showpacs,preprintnumbers,amsmath,amssymb]{revtex4}

\usepackage{graphicx}
\usepackage{dcolumn}
\usepackage{bm}
\usepackage{color}

\begin{document}

\preprint{APS/123-QED}

\title{ Magnetic measurements at pressures above 10 GPa in a miniature ceramic anvil cell for a superconducting quantum interference device magnetometer\footnote{Rev. Sci. Instrum. {\bf 83}, 053905 (2012).}}

\author{Naoyuki Tateiwa$^{1}$}
\email{tateiwa.naoyuki@jaea.go.jp} 
\author{Yoshinori Haga$^{1,2}$}%
\author{Tatsuma D Matsuda$^{1}$}%
\author{Zachary Fisk$^{1,3}$}

\affiliation{
$^{1}$Advanced Science Research Center, Japan Atomic Energy Agency, Tokai, Naka, Ibaraki 319-1195, Japan\\
$^{2}$JST, Transformative Research-Project on Iron Pnictides (TRIP), Tokyo 102-0075, Japan\\
$^{3}$University of California, Irvine, California 92697, USA\\
}
\date{\today}

\begin{abstract}
 A miniature ceramic anvil high pressure cell (mCAC) was earlier designed by us for magnetic measurements at pressures up to 7.6 GPa in a commercial superconducting quantum interference (SQUID) magnetometer [N. Tateiwa {\it et al.,} Rev. Sci. Instrum. {\bf 82}, 053906 (2011)]. Here, we describe methods to generate pressures above 10 GPa in the mCAC. The efficiency of the pressure generation is sharply improved when the Cu-Be gasket is sufficiently preindented. The maximum pressure for the 0.6 mm culet anvils is 12.6 GPa when the Cu-Be gasket is preindented from the initial thickness of 300 to 60 $\mu$m.  The 0.5 mm culet anvils were also tested with a rhenium gasket. The maximum pressure attainable in the mCAC is about 13 GPa. The present cell was used to study YbCu$_2$Si$_2$ which shows a pressure induced transition from the non-magnetic to magnetic phases at 8 GPa. We confirm a ferromagnetic transition from the dc magnetization measurement at high pressure. The mCAC can detect the ferromagnetic ordered state whose spontaneous magnetic moment is smaller than 1 ${\mu}_{\rm B}$ per unit cell. The high sensitivity for magnetic measurements in the mCAC may result from the the simplicity of cell structure. The present study shows the availability of the mCAC for precise magnetic measurements at pressures above 10 GPa.
\end{abstract}

\maketitle

\section{Introduction}

  The application of pressure to materials provides a powerful method of tuning various physical properties to search for new phase transitions in the strongly correlated electron systems. Interesting physical phenomena such as superconductivity have been found near the boundary to a magnetic phase\cite{buzea}. Various experimental techniques have progressed for the measurement of physical quantities such as electrical resistivity, heat capacity and magnetization under high pressure\cite{eremets}. Magnetization is a fundamental physical property characterizing the response of a material to applied magnetic field. Magnetic measurements under high pressure are essential for the study of pressure-induced physical phenomena.   
  
  Several types of high pressure cells have been made for magnetic measurements at high pressure in a commercial superconducting quantum interference (SQUID) magnetometer where measurements can be done under automatic control of temperature and magnetic field~\cite{reich,kamarad,mito1,mito2,alireza,giriat,kobayashi}.  The magnetometer can resolve magnetic moment changes as small as 10$^{-8}$ emu. The pressure cells for the magnetometer can be divided into two categories, piston-cylinder and opposed anvil cells such as the diamond anvil cell (DAC).  Most of the cells for the SQUID magnetometer are of piston-cylinder type~\cite{reich,kamarad}. Large sample volume is one of the key advantages of this cell. A precise magnetization measurement is possible. However, the pressure range is limited to at most 1.5 GPa since the inner bore diameter of the SQUID magnetometer is only 9 mm. The DAC for the magnetometer can generate high pressures up to 15 GPa~\cite{mito1,alireza,giriat}. However, the volume of the sample space in the DAC is less than 0.01 mm$^3$. It can be used for the ferromagnetic or superconducting compounds with large absolute magnetization. Specialized techniques are necessary to use the DAC and the production cost is high ($\sim$ 10$^4$ $\$$). An easier and more inexpensive method should be developed. Magnetization measurements at high pressure are possible up to 3.0 GPa using an indenter type cell~\cite{kobayashi}. The large background magnetization from the Ni-Cr-Al gasket is a serious problem. The gasket is pressed only by the cone part of the anvil and the mechanical support for the sample space is not enough.  The sample space in the gasket deforms radially under compression above the tensile strength of the Ni-Cr-Al alloy ($\sim$ 2.3 GPa), which produces two features in the indenter cell. One is the decrease in the efficiency of pressure generation at higher pressures and the other is the development of uniaxial (deviatoric) stress in the sample space above the solidification pressure of the pressure-transmitting medium. The strength of the uniaxial stress correlates with the maximum shear stress of the pressure-transmitting medium. The latter feature causes difficulties that the pressure effects on the electronic state differ depending on pressure-medium used in each experiment~\cite{kotegawa1,kotegawa2}. It is important to suppress the radial deformation for generating higher pressures with good quality. The deformation can be avoided when the gasket is pressed by the surface of the anvils in the opposed anvil cell.

   Recently, we have proposed a miniature opposed-anvil high-pressure cell for use with the SQUID magnetometer\cite{tateiwa1,tateiwa2}. The anvils are made of inexpensive composite ceramic (FCY20A, Fuji Die Co.). This cell is abbreviated here as mCAC. The simplified mCAC without anvil alignment mechanism is easy-to-use for researchers who are not familiar with high-pressure technology. The production cost is about one tenth of that of the DAC. The background magnetization in the mCAC is far smaller than that in the indenter cell. The cell can be used for antiferromagnetic compounds with smaller magnetization. The maximum pressure was 7.6 GPa when the 0.6 mm culet anvils were used. In this study, we have developed methods to generate pressures above 10 GPa in the mCAC.

    \begin{figure}[]
\includegraphics[width=8cm]{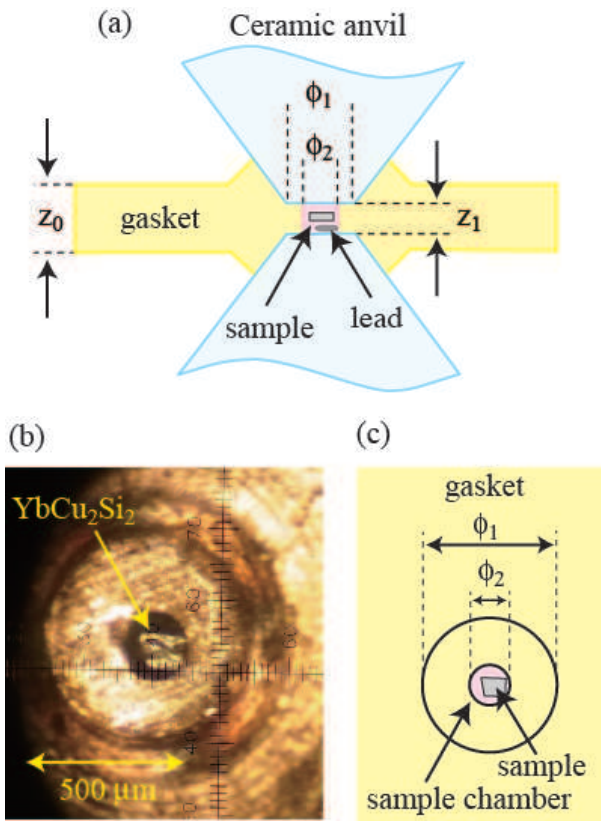}
\caption{\label{fig:epsart}(Color online)(a)Schematic illustration of the gasket and ceramic anvils in the mCAC, (b)Photograph of the Cu-Be gasket after the preindentation using a pair of the ${\phi}_1$ = 0.6 mm anvils, and (c) schematic illustration. Sample (YbCu$_2$Si$_2$) and pressure manometer Pb (behind the sample) are loaded in the sample chamber whose diameter is ${\phi}_2$ = 0.2 mm. A scale in the microscope corresponds to 20 $\mu$m.}
\end{figure}

 \section{EXPERIMENTAL METHODS}    
 We have used the mCAC with the 0.6 and 0.5 mm culet anvils for the present study. In particular, the effect of the preindentation of the Cu-Be gasket has been carefully studied. Figure 1 and Table 1 show the experimental configuration and conditions of the present study. We have tested three Cu-Be gaskets with initial thickness of 0.30 mm. The Cu-Be gaskets were preindented to a thickness of 0.06 and 0.15 mm in Run 1 and 2, respectively. In Run 3, the gasket was not preindented. The rhenium (Re) gasket was also tested with the 0.5 mm culet anvils (Run 4). The rhenium gasket is used without preindentation. In Run 1 to 4, a lead (Pb) pressure manometer was loaded in the sample chamber filled with the pressure-transmitting medium glycerin\cite{tateiwa3}. For the magnetization measurement on YbCu$_2$Si$_2$, the sample and Pb were loaded in the chamber. Figure 1 (b) shows a photograph of the Cu-Be gasket after the preindentation using a pair of the ${\phi}_1$ = 0.6 mm anvils. A scale in the microscope corresponds to 20 $\mu$m. A sample (YbCu$_2$Si$_2$) and pressure manometer Pb (behind the sample) were loaded in the sample chamber whose diameter was ${\phi}_2$ = 0.2 mm.
 
  The magnetization measurement has been done using the SQUID magnetometer MPMS from Quantum Design (USA)\cite{mpms}. To obtain the magnetization of a sample, the SQUID response of the pressure cell is collected with and without the sample and the difference signal is fitted to a calculated form assuming a point dipole moment using a specially written external program as described in the reference 12. The pressure values at low temperatures were determined by the pressure dependence of the superconducting transition temperature in lead\cite{smith,eiling,wittig}.

         \begin{table*}[t]
\caption{\label{tab:table1}%
Experimental conditions. ${\phi}_1$ : culet size of the anvil. ${\phi}_2$: diameter of the sample space, $z_0$: initial thickness of the gasket,  $z_1$: thickness of the gasket after preindentation, and $P_{max}$ : maximum pressure. }
\begin{ruledtabular}
\begin{tabular}{cccccccc}
\textrm{Run}&
\textrm{gasket}&
\textrm{${\phi}_1$ (mm)}&
\textrm{${\phi}_2$ (mm)}&
\textrm{$z_0$ (mm)}&
\textrm{$z_1$ (mm)}&
\textrm{$P_{max}$ (GPa)}\\
\colrule
1 &Cu-Be &0.60& 0.30 &0.30& 0.06 & 12.6 $\pm$ 0.5\\
2 &Cu-Be &0.60& 0.30 &0.30& 0.15 & 7.6 $\pm$ 0.4 \\
3 &Cu-Be  &0.60& 0.30 &0.30& 0.30 & 5.0 $\pm$ 0.3 \\
4 & Rhenium  &0.50&0.20&0.05& without preindentation & 11.5 $\pm$ 0.3 \\
\end{tabular}
\end{ruledtabular}
 \end{table*}
 
\section{RESULTS AND DISCUSSIONS}
\subsection{Generation of high pressure above 10 GPa}
     Figure 2 shows the temperature dependence of the magnetization of Pb at 6.9, 9.1, 11.6, and 12.6 GPa in magnetic field of 10 Oe with the 0.6 culet anvils (Run 1). The data at 9.1, 11.6, and 12.6 GPa are shifted along the vertical axis for clarity.  At low temperatures, a large drop of the magnetization associated with the Meissner effect of the superconducting transition was observed. The value of $T_{sc}$ was determined from the peak temperature in the temperature derivative of magnetization ${\partial M}/{\partial T}$ shown as arrows in the figure. The magnetization becomes negligibly small above $T_{sc}$. The superconducting transition temperature $T_{sc}$ of Pb is 7.19 K at ambient pressure. The value of $T_{sc}$ is 4.9 K for an applied load of 220 kgf. This indicates that the pressure inside the sample chamber is 6.9 GPa. For the applied load of 333 kgf , the superconducting transition temperature is 3.73 K which corresponds to 12.6 GPa. 
   \begin{figure}[t]
\includegraphics[width=8cm]{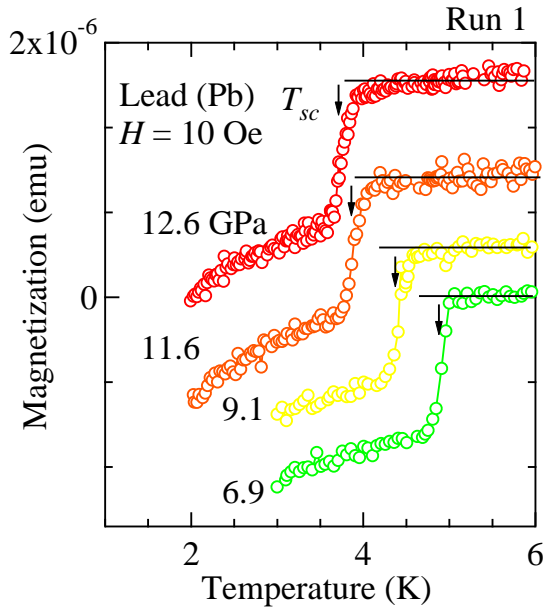}
\caption{\label{fig:epsart}(Color online)Temperature dependence of the magnetization of Pb at 6.9, 9.1, 11.6, and 12.6 GPa in magnetic field of 10 Oe with the 0.6 culet anvils (Run 1). The data at 9.1, 11.6, and 12.6 GPa are shifted along the vertical axis for clarity.}
\end{figure} 

    \begin{figure}[t]
\includegraphics[width=8cm]{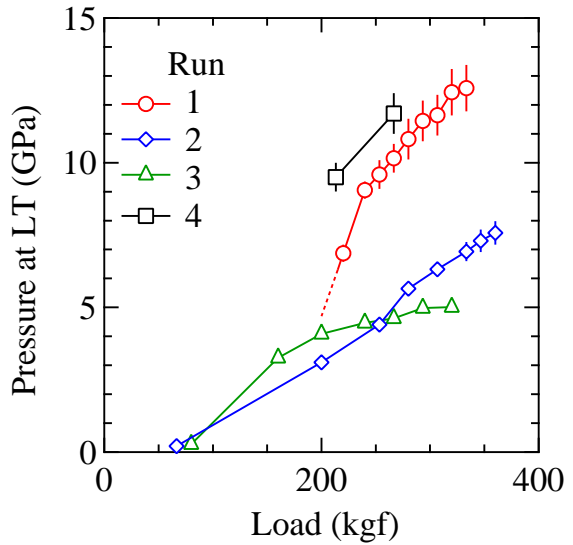}
\caption{\label{fig:epsart}(Color online) Relations between the applied load at room temperature and the pressure values at low temperatures for Run 1, 2, 3, and 4.}
\end{figure} 

    \begin{figure}[t]
\includegraphics[width=8cm]{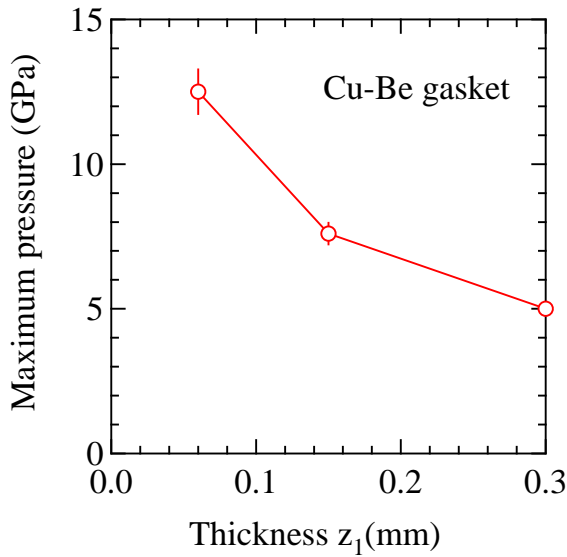}
\caption{\label{fig:epsart}(Color online) Relation between the thickness of the Cu-Be gasket after the preindentation $z_1$ and the maximum pressure.}
\end{figure}

   Figure 3 shows the relation between the applied load at room temperature and the pressure value at low temperature for the Run 1, 2, 3, and 4. The maximum pressure $P_{max}$ depends on the thickness $z_1$ after the preindentation when the Cu-Be gaskets were used. The pressure value shows saturation with increasing applied load when the gasket is not preindented (Run 3). The value of $P_{max}$ is at most 5.0 GPa. The maximum pressure is sensitive to the thickness of the Cu-Be gasket $z_1$ after the preindentation as shown in Figure 4. It seems that the gasket should be preindented to less than 0.1 mm from the initial thickness of 0.30 mm in order to generate high pressures above 10 GPa.
 
 We have tested the 0.5 mm culet anvils with the preindented Cu-Be gasket. The sample chamber deforms non-symmetrically and the gasket was broken at higher pressure.  The deformation may be due to misregistrations between center positions in the two anvils and the Cu-Be gasket. There is no anvil-alignment mechanism in the mCAC as mentioned before. Next, we tested the rhenium gasket with the 0.5 mm culet anvils. The tensile strength of rhenium is about two times larger than that of the Cu-Be alloy. Therefore, we expected the generation of the higher pressure. The relation between the applied load and the pressure was shown in Fig. 3 (Run 4).  The pressure value increases from 9.5 GPa at 213 kgf to 11.7 GPa at 267 kgf. On further increasing the load, the anvils were broken at 293 kgf. This load corresponds to 12.9 GPa if the relation between the load and the pressure value in Run 4 is extrapolated linearly. The present study suggests that the maximum pressure attainable with the mCAC without the anvil alignment mechanism is about 13 GPa. The anvil alignment mechanism is required for the small culet anvils to generate higher pressure. Future study is necessary for the maximum pressure generated by anvils made of the present composite ceramic (FCY20A).
 
  We note the influence of the superconducting transition on the rhenium gasket to the magnetic measurements. The contribution from the superconducting transition in the rhenium gasket to the magnetic data is dominant since the volume of the gasket is far larger than those of the pressure manometer Pb and the sample. The rhenium gasket is not appropriate for magnetic measurements at low magnetic fields and low temperatures. The superconducting transition temperatures at 9.5 and 11.7 GPa under magnetic field of 10 Oe are 3.5 and 4.0 K, respectively.  The superconducting transition temperature in rhenium metal was reported to be 1.699 K at ambient pressure\cite{chu}. The pressure dependence of the transition temperature is reported only up to 1.8 GPa. The present result suggests that the transition temperature increases with increasing pressure. However, it is not clear whether the present pressure change of the transition temperature reflects the effect of the hydrostatic pressure since the rhenium gasket is directly pressured by anvils. It was reported that the electronic state of rhenium metal is sensitive to the uniaxial strain\cite{antonov}.   
   
    \begin{figure}[t]
\includegraphics[width=8cm]{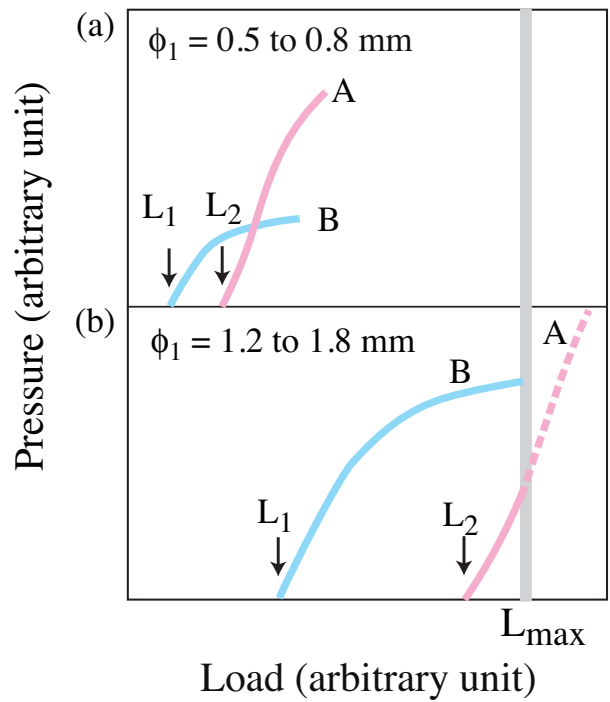}
\caption{\label{fig:epsart}(Color online)Schematic illustration of the relations between the applied load and the pressure value for (a) the 0.5-0.8 mm anvils and (b) the 1.2-1.8 mm anvils. Lines A and B indicate the relations for the Cu-Be gaskets with and without preindentation, respectively. $L_1$ and $L_2$ are threshold values of the applied load where the pressure value starts to increase. $L_{max}$ is the maximum load limit for the present ceramic anvil estimated as 1100 kgf. }
\end{figure}  

  The present study shows the importance of the preindentation of the Cu-Be gasket when the ${\phi}_1$ = 0.6 culet anvils are used. On the other hand, it is not better to preindent the Cu-Be gasket in the ceramic anvil cell when the ${\phi}_1$ = 1.2 to 1.8 mm culet anvils are used.  Figure 5 (a) and (b) show the schematic illustration of the relations between the applied load and the pressure value for (a) the ${\phi}_1$  = 0.5-0.8 mm anvils and (b) the 1.2-1.8 mm anvils. Lines A and B represent the relations for the gaskets with and without the preindentation, respectively. Firstly, we discuss the case for (a). The pressure value for the gasket without the preindentation starts to increase above a threshold value $L_1$ of the applied load and saturate in the lower load (Line B). For the gasket with the preindentation, The threshold value $L_2$ of the load is a few times larger than $L_1$ and the higher pressure can be generated (Line A). The threshold values $L_1$ and $L_2$ are far smaller than the maximum load limit $L_{max}$ for the present ceramic anvil estimated as 1100 kgf. The preindentation for the CuBe gasket is important.  The relations shown in Fig. 5 (a) are experimentally demonstrated in Fig. 3. Meanwhile, the value of $L_2$ for the 1.2-1.8 mm culet anvils becomes close to $L_{max}$ as shown in Fig. 5 (b). The higher pressures cannot be generated below $L_{max}$ with the preindented gasket. The applied load larger than $L_{max}$ is required to achieve higher pressure than that using the gasket without the preindentaion (Dotted line A in Fig. 5 (b)). It is better to use the Cu-Be gasket without the preindentation. For the rhenium gasket, we did not preindent it in this study since the rhenium metal is harder than the Cu-Be alloy and the pressure value does not show saturation in the pressure region up to 12 GPa.

\subsection{Measurement on YbCu$_2$Si$_2$}
 We illustrate the performance of the present methods described in this paper with a study YbCu$_2$Si$_2$. YbCu$_2$Si$_2$ crystallizes into the tetragonal ThCr$_2$Si$_2$-type structure. This is a paramagnetic compound with a moderately high value of the linear specific heat coefficient ${\gamma}{\,}{\simeq}$ 135 mJK$^{-2}$mol${^{-1}}$\cite{dung}. Previous high pressure studies suggested a pressure-induced magnetic phase above 8 GPa\cite{yadri,colombier,winkelmann}. The ac-magnetic susceptibility ${\chi}_{ac}$ and calorimetry $C_{ac}$ under magnetic field suggested a ferromagnetic transition\cite{fernandez}. It is necessary then to detect the ferromagnetic component from the dc magnetic measurement under high pressure. We have measured the magnetization in YbCu$_2$Si$_2$ under high pressure with the mCAC using the 0.6 mm cult anvils. The Cu-Be gasket was preindented to $z_1$ = 0.08 mm from the initial thickness of $z_0$ = 0.30 mm. A high quality single crystal sample was used in this measurement and the detail of the sample preparation is given in the reference 21. The size of the single crystal sample was 0.11 $\times$ 0.09 $\times$ 0.03 mm$^3$. The sample and the Pb pressure manometer were placed in the sample space filled with the pressure-transmitting medium glycerin as shown in Fig. 1 (b). The magnetic field $H$ was applied parallel to the magnetic easy axis (the [001] direction) of the tetragonal crystal structure.
       \begin{figure}[t]
\includegraphics[width=8cm]{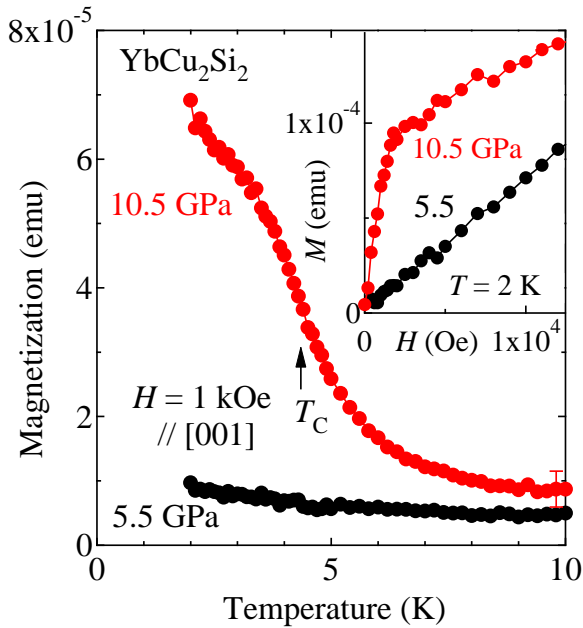}
\caption{\label{fig:epsart}(Color online)Temperature dependence of the magnetization under magnetic field of 1 kOe at 5.5 and 10.5 GPa.  The magnetic field $H$ was applied parallel to the magnetic easy axis (the [001] direction ). The inset shows the magnetization curves measured at 2 K for the two pressures.
}
\end{figure}

  Figure 6 shows the temperature dependence of the magnetization under magnetic field of 1 kOe at 5.5 and 10.5 GPa. The magnetization at 5.5 GPa shows a monotonous temperature dependence.  Meanwhile, the magnetization at 10.5 GPa increases strongly with decreasing temperature below 5 K, suggesting a ferromagnetic ordering. The inset of Fig. 6 shows the magnetization curves for the two pressures measured at 2 K. The magnetization increases monotonously with increasing magnetic field at 10.5 GPa. The magnetization at 10.5 GPa shows a typical ferromagnetic behavior.  It is concluded from these results that the pressure-induced magnetic transition in YbCu$_2$Si$_2$ is ferromagnetic. The ferromagnetic transition temperature $T_{\rm C}$ at 1 kOe was estimated as 4.3 K $\pm$ 0.2 K from the peak position in the temperature dependence of the derivative of the magnetization ${\partial M}/{\partial T}$. The transition temperature determined from the magnetization at 100 Oe (data not shown) was 4.4 $\pm$ 0.3 K. These values are roughly consistent with the previous studies\cite{yadri,colombier,winkelmann,fernandez}. 
      
 The previous study showed an electronic phase separation of the ferromagnetic and paramagnetic states above 8 GPa in YbCu$_2$Si$_2$\cite{winkelmann}. The M\"{o}ssbauer spectra were fitted by a superposition of a magnetic and nonmagnetic component and the value of the magnetic moment was estimated as 1.25 ${\mu}_{\rm B}$/Yb from the magnetic subspectrum at 8.9 GPa.  The observed spontaneous magnetic moment should be smaller than the value in the macroscopic magnetic measurement. Indeed, the spontaneous magnetic moment was estimated as 0.36 $\pm$ 0.05 ${\mu}_{\rm B}$/Yb at 2 K and 10.5 GPa in this experiment.

 Finally, we compare the performance of the mCAC with that of the DAC. The previous studies have reported the observation of ferromagnetic transitions in several materials with the DAC\cite{mito1,mito2,alireza,giriat}. The values of the spontaneous magnetic moments in the materials are generally larger than 1 ${\mu}_{\rm B}$ per unit cell. Meanwhile, the present study shows that the mCAC can detect the ferromagnetic ordered state around 10 GPa whose spontaneous magnetic moment is significantly less than 1.0 ${\mu}_{\rm B}$ per unit cell.  This high sensitivity for magnetic measurements may result from the the simplicity of cell structure. In general, the pressure cells with complicated structure often generate larger artificial inductive voltage in the pick-up coils, leading to a lower signal to noise ratio.

\section{CONCLUSION}
 
 In summary, we have developed methods to generate pressures above 10 GPa in a miniature ceramic anvil high pressure cell (mCAC). The efficiency of the pressure generation is sharply improved when the Cu-Be gasket is sufficiently preindented. The maximum pressure for the 0.6 mm culet anvils is 12.6 GPa. The 0.5 mm culet anvils were also tested with a rhenium gasket. The maximum pressure  attainable in the mCAC is approximately 13 GPa. We have applied the present methods to the study of the pressure-induced magnetic phase in YbCu$_2$Si$_2$. The ferromagnetic ordering state is confirmed by the dc magnetic measurements at high pressure. The mCAC can detect the ferromagnetic ordered state whose spontaneous magnetic moment is significantly less than 1.0 ${\mu}_{\rm B}$ per unit cell. The present study shows the availability of the mCAC for magnetic measurements at pressures across 10 GPa in the study of the strongly correlated electron system.

   \section{ACKNOWLEDGMENTS}
    This work was supported by a Grant-in-Aid for Scientific Research on Innovative Areas ``Heavy Electrons (Nos. 20102002 and 23102726), Scientific Research S (No. 20224015), A(No. 23246174), and C (No. 22540378), and for Young Scientists (B) (No. 2274021) from the Ministry of Education, Culture, Sports, Science and Technology (MEXT) and Japan Society of the Promotion of Science (JSPS).

\bibliography{apssamp}

\begin{references}

\bibitem{buzea}C. Buzea and K. Robble, Supercond. Sci. Technol. {\bf 18}, R1 (2005). 

\bibitem{eremets}M. I. Eremets,  {\it High-pressure Experimental Methods} (Oxford University Press, Oxford, 1996). 


\bibitem{reich}S. Reich and T. Godin, Meas. Sci. Technol. {\bf 7}, 1079 (1996).

\bibitem{kamarad}J. Kamar{\'a}d, Z. Mach{\'a}tov{\'a}, and Z. Arnold, Rev. Sci. Instrum. {\bf 75}, 5022 (2004).

\bibitem{mito1}M. Mito, M. Hitaka, T. Kawae, K. Takeda, T. Kitai, and N. Toyoshima, Jpn. J. Appl. Phys. {\bf 40}, 6641 (2001).
\bibitem{mito2}K. Takeda, and M. Mito, J. Phys. Soc. Jpn. {\bf 71}, 729 (2002).

\bibitem{alireza}P. L. Alireza and G. G. Lonzarich, Rev. Sci. Instrum. {\bf 80}, 023906 (2009).
\bibitem{giriat}G. Giriat, W. Wang, J. P. Attfield, A. D. Huxley, and K. V. Kamenev, Rev. Sci. Instrum. {\bf 81}, 073905 (2010).

\bibitem{kobayashi}T. C. Kobayashi, H. Hidaka, H. Kotegawa, K. Fujiwara, and M. I. Eremets, Rev. Sci. Instrum. {\bf 78}, 023909 (2007).

\bibitem{kotegawa1}H. Kotegawa, T. Kawazoe, H. Sugawara, K. Murata and T. Tou: J. Phys. Soc. Jpn. {\bf 78}, 0837002 (2009).
\bibitem{kotegawa2} H. Kotegawa, S. Araki, T. Akazawa, A. Hori, Y. Irie, S. Fukushima, H. Hidaka, T. C. Kobayashi, K. Takeda, Y. Ohishi, K. Murata, E. Yamamoto, S. Ikeda, Y. Haga, R. Settai, and Y. {\=O}nuki: Phys. Rev. B {\bf 84}, 054524 (2011).

\bibitem{tateiwa1}N. Tateiwa, Y. Haga, Z. Fisk, and Y. {\=O}nuki  Rev. Sci. Instrum. {\bf 82}, 053906 (2011).

\bibitem{tateiwa2}N. Tateiwa, and Y. Haga, Japanese Patent Tokugan No. 2011-054153 (pending).


\bibitem{tateiwa3}N. Tateiwa and Y. Haga,  Rev. Sci. Instrum. {\bf 80}, 123901 (2009).

\bibitem{mpms}Quantum Design Co., web site: http://www.qdusa.com/

\bibitem{smith} T. F. Smith, C. W. Chu, and M. B. Maple, Cryogenics {\bf 9}, 53 (1969).
\bibitem{eiling} A. Eiling and J. S. Schilling, J. Phys. F : Metal Phys., {\bf 11}, 623 (1981).
\bibitem{wittig} B. Bireckoven and J. Wittig, J. Phys. E: Sci. Instum.  {\bf 21}, 841 (1988).

\bibitem{chu} C. W. Chu, T. F. Smith, and W. E. Gardner, Phys. Rev. Lett.  {\bf 20}, 198 (1968).
\bibitem{antonov} V. I. Smelyansky, A. Ya. Perlov, and V. N. Antonov, J. Phys. Condens. Matter {\bf 3} 9033 (1991).

\bibitem{dung}N. D. Dung, T. D. Matsuda, Y. Haga, S. Ikeda, E. Yamamoto, T. Ishikura, T. Endo, S. Tatsuoka, Y. Aoki, H. Sato, T. Takeuchi, R. Settai, H. Harima, and Y. {\=O}nuki, J. Phys. Soc. Jpn. {\bf 78}, 084711 (2009).

\bibitem{yadri}K. Alami-Yadri and D. Jaccard, Eur. Phys. J. B {\bf 6}, 5 (1998).
\bibitem{colombier}E. Colombier, D. Braithwaite, G. Lapertot, B. Salce, and G. Knebel, Phys. Rev. B {\bf 79}, 245113 (2009).
\bibitem{winkelmann}H. Winkelmann, M. M. Abd-Elmeguid, H. Micklitz, J. P. Sanchez, P. Vulliet, K. Alami-Yadri, and D. Jaccard, Phys. Rev. B {\bf 60}, 3324 (1999).
\bibitem{fernandez} A. Fernandez-Pa{\~{n}}ella, D. Braithwaite, B. Salce, G. Lapertot, and J. Flouquet, Phys. Rev. B {\bf 84}, 134416 (2011).


\end{references}

 \newpage

\end{document}